# Greater benefits of deep learning-based computer-aided detection systems for finding small signals in 3D volumetric medical images

Devi S. Klein[1,1], Srijita Karmakar[1], Aditya Jonnalagadda[2], Craig K. Abbey[1] & Miguel P. Eckstein[1,2,3]

Department of Psychological and Brain Sciences[1]

Department of Electrical and Computer Engineering[2]

Department of Computer Science[3]

University of California, Santa Barbara

Santa Barbara, CA, 93106 United States

---

[1] Devi Klein is the corresponding author (email: dklein@ucsb.edu).

# Abstract

## Purpose

Radiologists are tasked with visually scrutinizing large amounts of data produced by 3D volumetric imaging modalities. Small signals can go unnoticed during the 3D search because they are hard to detect in the visual periphery. Recent advances in machine learning and computer vision have led to effective computer-aided detection (CADe) support systems with the potential to mitigate perceptual errors.

## Approach

Sixteen non-expert observers searched through digital breast tomosynthesis (DBT) phantoms and single cross-sectional slices of the DBT phantoms. The 3D/2D searches occurred with and without a convolutional neural network (CNN)-based CADe support system. The model provided observers with bounding boxes superimposed on the image stimuli while they looked for a small microcalcification signal and a large mass signal. Eye gaze positions were recorded and correlated with changes in the area under the ROC curve (AUC).

## Results

The CNN-CADe improved the 3D search for the small microcalcification signal ($\Delta AUC = 0.098, p = 0.0002$) and the 2D search for the large mass signal ($\Delta AUC = 0.076, p = 0.002$). The CNN-CADe benefit in 3D for the small signal was markedly greater than in 2D ($\Delta\Delta AUC = 0.066, p = 0.035$). Analysis of individual differences suggests that those who explored the least with eye movements benefited the most from the CNN-CADe ($r = -0.528, p = 0.036$). However, for the large signal, the 2D benefit was not significantly greater than the 3D benefit ($\Delta\Delta AUC = 0.033, p = 0.133$).

## Conclusion

The CNN-CADe brings unique performance benefits to the 3D (vs. 2D) search of small signals by reducing errors caused by the under-exploration of the volumetric data.

# 1. INTRODUCTION

Digital breast tomosynthesis (DBT) is becoming the standard imaging modality for early cancer screening within the United States [1]. DBT affords a quasi-3D rendering of the patient's anatomy that reduces tissue superposition and signal occlusion inherent in the 2D planar views generated from digital mammography image reconstruction algorithms [2]. Radiologists interpret DBT images by freely scrolling back and forth through cross-sectional slices of the volumetric data—displayed one at a time on a computer monitor—to visually segment masses, microcalcifications, and architectural distortions from surrounding parenchyma [3].

3D volumetric images, however, pose new challenges to the radiological decision-making process [4] because of the increased data requiring visual scrutiny. It would be prohibitively time-consuming to scan exhaustively, with eye movements, each cross-sectional slice in the stack of images before terminating one's search. Therefore, radiologists must adopt new search strategies to perform 3D visual searches [5–7]. For example, a recent eye-tracking study demonstrated that radiologists and trained human observers rely on peripheral vision when scrolling through 3D volumetric images. Due to under-exploring the 3D image stack with eye movements, trained observers and radiologists miss small signals that are hard to detect in the visual periphery [8,9]. Specifically, under-exploration leads to search errors of small signals, a miss that occurs because the observer failed to direct their center of gaze to the signal's location [10,11].

Recent advances in deep learning-based computer-aided detection (CADe) algorithms provide a promising avenue for mitigating search errors in 3D volumetric images. First, unlike human observers, Convolutional Neural Network (CNN)-based CADe systems are not constrained by attentional bottlenecks that are a consequence of foveated vision—high spatial acuity in the fovea and low spatial acuity in the peripheral visual field [12]. The convolution kernels in a CNN can process each voxel in a large 3D volumetric image in parallel while simultaneously filtering for both high and low spatial frequency information [13]. Second, CNN-based CAD algorithms—models that can perform both classification (CADx) and detection simultaneously—have obtained non-inferior performance relative to expert radiologists [14,15]. Thus, these artificial intelligence-based support systems can work in parallel with an attending radiologist as a "co-pilot" to enhance and augment their workflow [16,17].

To date, no systematic vision science investigation delineates how a CNN-CADe algorithm benefits visual search in 2D versus 3D imaging modalities. For instance, does the CNN-CADe induce different performance benefits for 2D and 3D imaging modalities, and do these benefits depend on whether the signal is spatially large or small? Moreover, what types of errors does the CADe system mitigate in 2D? Are they the same types of errors as in 3D? Do individuals who under-explore the image/volume with eye movements benefit the most from the additional information provided by the CNN-CADe adjunct?

To answer these questions, we conduct an eye-tracking study to evaluate the utility of a CNN-CADe support system on human detection performance. The model produces bounding boxes on suspicious locations made viewable to naive (trained) observers while they perform a visual search task for simulated cancers in DBT phantoms (50% prevalence rate). Specifically, trained observers search with (and without) the CNN-CADe for a small microcalcification-like signal and a large mass-like signal embedded in 3D breast phantoms (3D search) and single slices of the phantoms (2D search).

We hypothesize that the CNN-CADe will guide an observer's eye movements to suspicious locations in the 3D volumetric image that would have otherwise been missed without it. We predict a more considerable reduction in microcalcification search errors in 3D than in 2D because it is relatively easy to explore most regions of a 2D image with eye movements in a time-efficient manner. For the large mass-like signal, we hypothesize that the search with the CNN-CADe will result in a less pronounced reduction in search errors in 3D because the mass is more detectable in the visual periphery than the small microcalcification-like signal. However, the mass-like signal is more difficult to recognize in 2D than 3D. Thus, we predict that the CNN-CADe will mitigate 2D recognition errors—misses that occur even after fixating the signal [18]. Finally, we hypothesize that observers with the highest degree of under-exploration of 3D images will benefit the most from the CNN-CADe adjunct when searching for a small microcalcification-like signal. To quantify an observer's personalized, effective exploration of 3D volumes, we combine their eye movement scan path data with an estimated Useful Field of View acquired from a separate task that measures an observer's peripheral detectability for each signal.

## 2. METHODS

### 2.1 Participants

Sixteen undergraduate students (62.5% female, age range 18-22) from the University of California, Santa Barbara, participated in this experiment for course credit. All participants provided informed written consent and were treated according to human subject research protocols approved by the University of California, Santa Barbara (protocol # - 12-23-0301). Participants maintained normal or corrected-to-normal vision throughout the duration of the experiment.

### 2.2. Apparatus

#### 2.2.1. Display Monitor

Participants interacted with stimuli on a medical grade grayscale DICOM monitor (Brand-Barco, type-MDNG-6121; 24 Hz refresh rate; 5.8 MP or 2096x2800 pixel resolution or 325x430 mm screen size) at a viewing distance of 750 mm in a darkened room (ambient luminance = 2 lux). 45 pixels on the monitor screen subtended 1 degree of visual angle (dva). We calibrated the monitor with a Barco LCD sensor (42630), and it passed a MediCal QAWeb DICOM GSDF compliance test with a maximum error of 7%.

#### 2.2.2. Eye-tracker

While participants engaged with the task, an eye tracker (SR Research Eyelink Desktop Mount) monitored their gaze position at 2000 Hz. Participants encountered a calibration and validation procedure at the beginning of each session and could recalibrate between trials if necessary. Each procedure used a nine-point grid, and successful calibration was met if the average validation error across the 9 grid points was less than 1 dva and the max error was less than 1.5 dva. Fixations and saccades were analyzed offline using the standard velocity and acceleration thresholds of 30 deg/s and 9,500 deg/s$^2$, respectively.

#### 2.2.3. Experiment control

The experiment used the Python package PsychoPy [19]. Events such as mouse scrolls and clicks were sampled at the monitor refresh rate of 24 Hz but synced to a wall clock via the ioHub event monitoring module in Psychopy to facilitate co-registration in the timing of these events with saccade and fixation data acquired from the eye tracker.

### 2.3. Stimuli

#### 2.3.1. Phantoms

Participants viewed anthropomorphic DBT phantoms that simulate the spatial arrangement of anatomical tissues (skin, Cooper's ligaments, adipose, and glandular) and lesions (microcalcifications and masses). The phantoms were generated with the OpenVCT virtual breast imaging tool from the University of Pennsylvania [20–22] using clinical acquisition geometry and clinical automatic exposure control settings (Selenia Dimensions, Hologic, Marlborough, MA). The 700 ml simulated phantoms were compressed in the mediolateral direction at 6.33 mm thickness with glandular tissue prevalence of 15%-25%. The spatial reconstruction parameters were set to 100 $\mu$m in-plane resolution and 1 mm depth sampling (Briona Standard; Real Time Tomography, LLC, Villanova, PA), producing a 3D voxel array of size 2048x1792x64. Each voxel of a phantom was stored as an unsigned 16-bit integer. For display purposes, we windowed the volumetric images between 5066 and 16907 and then applied a linear rescaling to conform with the backend display functions in Psychopy, which requires 8-bit images. We utilized 160 unique 3D DBT phantoms for the search tasks described below.

### 2.3.2. Signals.

Participants searched for two types of simulated lesions. The first signal was a solid sphere (0.3 mm diameter, 0.06 dva in the xy plane) akin to a small microcalcification lesion and spanned ~6 cross-sectional slices. The second signal resembled a mass lesion and was generated with a combination of several 3D ellipsoids with an average diameter of 7 mm (0.5 dva in the x, y plane). The density of the mass lesion decreased towards the edges of the signal profile, causing it to blend in with the anatomical background to a greater extent than the microcalcification signal. The mass signal spanned ~15 cross-sectional slices. Both signals were added to the background before the windowing and rescaling operations described above.

## 2.4. Search task

### 2.4.1. Experimental design

Human observers performed a Yes/No localization task [23,24] and reported whether a single signal was present or absent in the image stimulus. The experiment had three within-subjects factors, each with two levels: imaging modality (2D and 3D), CNN-CADe (searching with and without CADe support), and signal type (microcalcification and mass), totaling eight conditions. The presentation order of the levels of the first two factors was counterbalanced across participants. For example, half of the participants started the experiment with the CNN-CADe support, followed by a washout period (minimum two weeks) before they saw the same stimuli without the CNN-CADe. Of those participants who completed the CNN-CADe conditions first, half searched through the 3D phantoms with the CADe before searching through the 2D slices of the phantoms with the CADe. The other half of the participants performed the 2D search before performing the 3D search. The same counterbalancing procedure between 2D and 3D searches was implemented for the other half of the participants who completed the search without the CNN-CADe before the washout period. The last factor, signal type, was combined into a single block of 160 trials (50% prevalence). In other words, one block contained 40 microcalcification-present trials, 40 microcalcification-absent trials, 40 mass-present trials, and 40 mass-absent trials. The presentation order was randomized across both signal type and ground truth status. Each block comprised 16 10-trial sessions with an enforced 2-minute break in between sessions to mitigate fatigue effects. Participants completed four blocks of trials, with the 2-week washout period occurring between blocks 2 and 3.

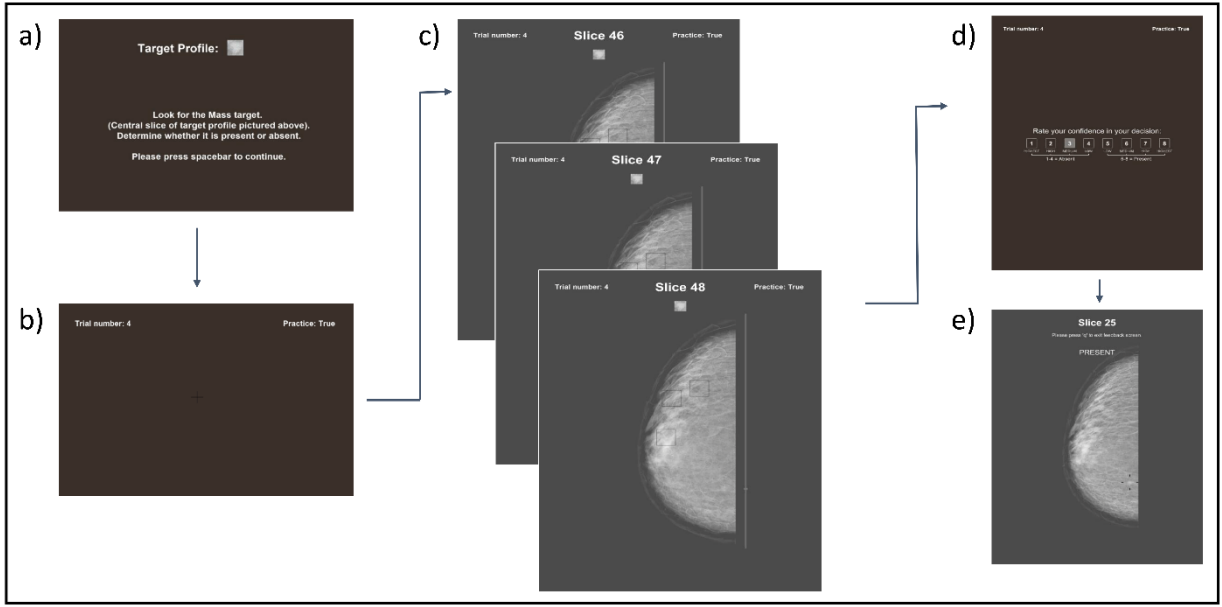

Figure 1. An example practice trial where a participant is looking for the mass signal in 3D, and they have access to the CNN-CADe output. a) At the beginning of each trial, the participant is informed of which signal they need to look for. In this case, it is the mass signal. A cropped image of the central slice of the mass signal is shown to the participant in addition to instructional text. b) Next, a black fixation cross is placed at a random (x, y) coordinate on top of a uniform-colored background. The participant needs to stare at the fixation cross for 1 second before proceeding to the next component of the trial. c) 3 of 64 slices of a DBT phantom are shown. When scrolling through all 64 slices, only one slice is presented on the monitor at a time. Three cues (square bounding boxes) are superimposed on top of the image stimulus to indicate where the CNN-CADe thinks the signal is located to the participant. Note that the cues persist across multiple adjacent slices at the same (x, y) coordinate. d) After viewing the image stimulus, the participant must input a rating (1-8) indicating their confidence that the signal was absent/present in the stimulus. e) During practice trials only, the participant is shown feedback at the very end of the trial. For the 3D search task, the DBT phantom slice containing the central slice of the signal is shown on signal-present trials. A black fiducial marker surrounds the signal's location to inform the participant where the signal is located in the (x, y) plane. The central slice number is also shown at the top of the display to inform the participant where the signal was centrally located in the third dimension.

The general structure of a trial for each of the four blocks is depicted in Figure 1. At the beginning of each trial, participants viewed a cropped 2D image of the signal they would need to search for (Figure 1a). Instructional text was also provided. The cropped image was taken from the central slice of the 3D signal profile, the cross-sectional slice corresponding to the signal's centroid. Next, they had to force-fixate a black cross placed at a random location on top of a gray background for 1 second to ensure the eye tracker was well-calibrated for the trial (Figure 1b). Afterward, the image stimulus was presented to the participants until they chose to terminate the search (Figure 1c). The interface of this trial component changed depending on which of the four blocks the participants were in. In the following sections, we provide a more complete description of what Figure 1c entailed for each of the four blocks. Upon completion of the search component of the trial, participants rated their confidence in their decision on a scale of 1-8 (Figure 1d). A rating of 1 corresponded to the highest confidence that the signal was absent, and a rating of 4 represented the lowest confidence that the signal was absent. Conversely, a rating of 5 corresponded to the lowest confidence in the signal's presence, and a rating of 8 corresponded to the highest confidence in the signal's presence.

### 2.4.2. 3D search without CNN-CADe (Block I)

Upon completion of the forced fixation component of a trial (Figure 1b), participants were shown the 3D image data. The "top" slice of the 3D DBT volume would always first appear on the monitor screen. Only one slice was shown to the participant at a time. The cropped image of the signal and a slice index tracker appeared at the top of the screen, above the DBT slice. This can be visualized in Figure 1c. The signal template reminded participants which signal they needed to look for, and the slice index tracker displayed the current slice they were on. To view each of the 64 slices that comprised the DBT phantom image data, participants could either manipulate the mouse scroll wheel or hold and drag a widget placed on a custom-designed scroll bar located on the right-hand side of the image stimulus (also shown in Figure 1c). The mouse scroll wheel allowed participants to scroll back and forth through adjacent slices. In contrast, the scroll bar provided the additional functionality of clicking on the bar to jump across multiple slices at a time.

Participants had unlimited time to perform the search and were instructed to click on the (x, y, z) coordinate that produced the most evidence of the signal's presence. They had to navigate to the slice with the highest signal contrast and click on the center of the signal profile. Clicking on the screen produced a circle (radius of 1 dva) for visual confirmation. They were instructed not to click on the screen if they did not see the signal. To end the trial, they pressed the space bar on the keyboard.

### 2.4.3. 3D search with CNN-CADe (Block II)

A 3D search trial with the CNN-CADe support available mirrored the 3D search task described above but with one crucial caveat. While scrolling through the cross-sectional slices of the 3D DBT data, a varying number of square bounding boxes (3.1 dva in width/length) would appear on the screen, overlaid on top of the DBT slices. An example of what the cue boxes looked like while the participants scrolled through 3 DBT slices is shown in Figure 1c. Participants were informed that the cued locations correspond to where a computer vision model predicts the signal may be located. The cue boxes on microcalcification-present and microcalcification-absent trials persisted across 5 slices (2 slices above and below the central slice on which the box was placed), and the cue boxes on mass trials persisted across 11 slices (5 above and below the center location). The choice for the cue box locations and the number of cue boxes that appeared on screen for a given trial is described in subsections 2.6.5 – 2.6.7. below.

### 2.4.4. 2D search without CNN-CADe (Block III)

Participants interacted with a single DBT slice while performing the 2D search task. Concerning Figure 1c, only one DBT slice would appear on the screen, and scrolling would be disabled. Furthermore, neither the slice index tracker nor the custom scroll bar was on the monitor screen. All other aspects of the search interface depicted in Figure 1c were held constant.

On signal-present trials, the DBT slice corresponding to the central slice of the signal was displayed. The image stimulus depicted in Figure 1e, sans the additional feedback text and markings, provides an example of what the participants saw when searching for the mass signal in 2D. The $32^{nd}$ slice of a signal-absent 3D DBT phantom was displayed to participants on signal-absent trials.

We used single DBT slices rather than simulated mammograms to model the search with a "2D imaging modality". The noise power spectrum of 2D mammograms differs from that of single-slice DBT images [25]. Therefore, an AI-based CADe system may have a differential impact on 2D mammograms versus single slices of DBT volumes because image acquisition parameters and postprocessing differ across these two modalities, which can cause differences in lesion conspicuity [26]. DBT slices allowed us to isolate differences in performance across 2D and 3D searches while controlling for confounds that the image generation process may introduce.

### 2.4.5. 2D search with CNN-CADe (Block IV)

The 2D search task with the CNN-CADe support replicated the 2D search task described above. However, it included cue boxes superimposed on top of the DBT phantom slice. Like in Block II, participants were informed that the cued locations represent the predicted locations made by the computer vision model for the mass/microcalcification signal.

### 2.4.6. Training and practice trials

Before completing the experiment blocks, participants partook in 4 practice blocks to familiarize themselves with the search tasks. For the 3D and 2D practice blocks without the CNN-CADe, there were 80 practice trials per block and 40 trials per signal (50% prevalence). Additionally, there were 20 practice trials (10 per signal and 50% prevalence) for the 2D and 3D search practice blocks with the CNN-CADe. These blocks were included so observers could estimate the CNN's performance and develop an internal model of incorporating its information into their decision-making process.

Block of practice trials were interleaved with each of the four experimental blocks. For example, those participants randomly assigned to the 2D CNN-CADe block completed the practice block without the CNN-CADe. This was done to help familiarize them with the overall task procedure and to develop an understanding of what the two signals looked like in a single DBT phantom slice. Then, they would complete the practice block with the CNN-CADe before starting the experimental block. Upon completion of the 2D CNN-CADe experimental block, those same participants would complete the 3D practice block without the CADe, followed by the 3D practice block with the CADe, before starting the 3D experimental block with the CNN-CADe. Without further training, these participants would continue to the last 2 experiment blocks after the washout period.

Practice trials maintained the same design as experiment trials. However, at the end of every practice trial, feedback was given to the participants. On signal-present 2D trials, a fiducial marker was superimposed on top of the DBT slice, centered on the signal's location. If participants made a localization click on the trial, the circle centered on where they clicked was also present on the screen so they could discern where they clicked relative to the signal's location. On signal-present 3D trials, the same DBT slices were shown as in 2D (i.e., the central slice of the signal in the DBT volume) but included the slice number on which the center of the 3D signal was placed. Figure 1e provides a graphical depiction of the feedback on a mass-present 3D trail. For signal-absent trials in 2D and 3D, a gray background with the text "ABSENT" was displayed to participants. Participants had unlimited time to review the feedback before proceeding to the next practice trial.

## 2.5. Microcalcification and mass peripheral detectability task

Upon completing the search tasks, participants partook in a forced fixation yes/no location-known-exactly detection task. We included this task to measure each participant's peripheral detectability of the microcalcification and mass signals [27–29]. Participants viewed 800 stimuli, 400 per signal. Each stimulus was a single slice of a DBT phantom sampled from a set of stimuli not shown to participants in the search task. Half the stimuli contained a signal (50% microcalcification and 50% mass), and the other half contained no signal. The detection tasks were segregated into two separate 400-trial blocks, one block per signal.

At the beginning of each trial in a block, participants stared at a black fixation cross, superimposed on a gray background, at the center of the computer monitor. A fiducial marker was also present on the screen. The marker was centered 5 dva from the center of the fixation cross, and it appeared at 1 of 4 polar angles on any given trial: 0, 90, 180, or 270 degrees. Participants were informed that the signal would appear in half of the trials at the cued location. Therefore, they needed to only covertly attend to the marker's location and ignore all other locations in the image stimulus. After staring at the fixation cross for 1 second, the image stimulus appeared on the screen for 200 ms. The trial would abort if participants attempted to make a saccade towards the cued location. Afterward, participants encountered the same rating scale as in the search tasks (Figure 1d). They had to indicate their confidence that the microcalcification/mass was present (or absent) at the cued location. In sum, we measured the peripheral detectability at 4 polar coordinates in the visual field (100 trials per coordinate and 50% prevalence).

Each block for measuring the extra-foveal processing of the microcalcification/mass signal was preceded by a block of practice trials. There were 16 trials, 2 per polar coordinate (1 signal-present trial and 1 signal-absent trial). Practice trials provided feedback at the end of each trial, similar to Figure 1e.

## 2.6. CNN-CADe

### 2.6.1. Model overview

Our study employed an encoder-decoder U-Net CNN architecture for image segmentation [30,31]. The encoder-decoder semantic segmentation architecture allowed us to preserve a one-to-one mapping between the input stimulus size and the model output size. For our experiment, the model output, being the probability of malignancy at each voxel/pixel location in the image, is a requisite for displaying cue boxes to human observers during the search task. Furthermore, we utilized nnU-net [32], an out-of-the-box segmentation tool built upon the basic U-Net architecture. nnU-net automates the preprocessing, network architecture, training, and post-processing configuration settings given domain-relevant information for the use case at hand (i.e., properties of the dataset, voxel spacing size, image modality, image size, etc. [33]). Moreover, nnU-net has outperformed specialized networks on various biomedical tumor segmentation tasks, demonstrating its generalizability to new datasets [32]. For this experiment, we trained and tested 4 models: microcalcification-2D, microcalcification-3D, mass-2D, and mass-3D.

### 2.6.2. Preprocessing

The input to the 3D models for training were the phantoms cropped to size 380x380x64 to improve training efficiency the non-cascade full-resolution network. The input to the 2D models for training were single slices of the phantoms of size 793x2048x1. We cropped the left-hand side of the image slices as it was a black background that provided no relevant information for training. The phantom resided on the right-hand side of the slices for all stimuli.

### 2.6.3. Network architecture.

The backbone of the U-Net architecture for each of the 4 separate models consisted of an encoder and decoder module. For the encoding stage, strided convolution was used to down-sample the input spatial dimensions while increasing the feature dimensionality. For the decoding stage, up-sampling was performed using transposed convolutions, thus gradually decreasing the feature dimensionality while increasing the spatial dimensions until the output matched the input dimension size. Both the encoder and decoder were comprised of two computational blocks. Within each block, convolution operations were followed by instance normalization and a leaky-ReLU nonlinearity operation. The nnU-Net utilized a Stochastic Gradient Decent optimization to minimize the cross-entropy and maximize the dice coefficient with a preconfigured learning rate and Nestrov momentum hyperparameters set to 0.01 and 0.09, respectively. A 'polyLR' (polynomial function) regime caused the learning rate to decay across training for each parameter group.

### 2.6.4. Training

We utilized 5-fold cross-validation for training (4 training sets and 1 validation set). Thus, a given model (e.g., microcalcification-2D) was an ensemble of 5 separate CNNs, which were later combined to make predictions in the test set. Each of the 5 constituent models was trained for 1000 epochs. One epoch for the microcalcification-3D or mass-3D model took 800 seconds to complete compared to 240 seconds for the microcalcification-2D or mass-2D models. The training was completed across 4 12 GB Nvidia GPUs. 500 cropped phantoms containing the microcalcification and 500 cropped phantoms containing the mass signal were utilized to train the 3D models. 1,500 single slices—3 slices per each of the 500 phantoms used in the 3D training set—were chosen for training the 2D models. The 3 slices per phantom corresponded to the central slice and the slices above and below the central slice.

| Model | AUC | Proportion of trials cue on signal location | Connected components parameters | | | Number of cues | | | |
| | | | | | | Signal-present trials | | Signal-absent trials | |
| | | | P(malignancy) threshold | Euclidean distance (x, y) pixels | Manhattan distance (z) slices | Mean | SD | Mean | SD |
| --- | --- | --- | --- | --- | --- | --- | --- | --- | --- |
| 3D calc | 1.0 | 0.775 | 0.9 | 350 | 7 | 5 | 1.961 | 3.9 | 1.736 |
| 2D calc | 0.931 | 0.775 | 0.2 | 350 | N/A | 1.1 | 0.304 | 0.825 | 0.549 |
| 3D mass | 0.743 | 0.775 | 0.8 | 160 | 11 | 7.875 | 2.221 | 8.675 | 3.133 |
| 2D mass | 0.696 | 0.775 | 0.1 | 140 | N/A | 1.150 | 0.662 | 1.375 | 0.807 |

Table 1. CNN-CAD model performance metrics and relevant parameters for computing the number of cues and their locations.

### 2.6.5. Post-processing

After training, a given ensemble model was fed the full-resolution 3D phantom or 2D slice shown to the trained human observers. The probability of malignancy score at each voxel (3D) or pixel (2D) location was binarized using a model-specific threshold. The P(malignancy) thresholds are shown in the third column of Table 1 for all 4 models. Voxels above the threshold were treated as signal (1), and voxels less than the threshold were treated as background (0). We then applied a connected components algorithm—26 connectivity for 3D binarized model outputs and 6 connectivity for 2D binarized outputs—to join contiguous/neighboring signal voxels into blobs [34]. This procedure resulted in multiple connected components of varying sizes per stimulus.

### 2.6.6. Testing

With the connected component output in hand, we chose the count of voxels/pixels comprising the largest component as the decision variable of the model for each of the 80 test stimuli per imaging modality and signal type combination. We assumed that the largest component would correspond to the actual signal location if present in the volume/image. Moreover, on average, phantoms containing a signal would have larger connected components than phantoms without a signal. Based on this decision variable, we computed the area under the receiver operating curve (AUC) for each of the 4 models to confirm their ability to discriminate signal from noise. The AUCs for each model can be found in the second column of Table 1.

### 2.6.7. Converting CNN output to CADe support tool

There are many candidate options for displaying the CNN output as a support tool to human observers. Previous studies have presented both a stimulus-level score and location-specific scores in the form of cue boxes [35,36], saliency maps [37], or "click-to-see" model probability scores for a specific location on the image (i.e., interactive decision support [15,38]). We converted the connected component output into cue boxes/prompts superimposed on the image stimulus. Our central hypothesis focused on microcalcification search errors in 3D. Interactive decision support would not mitigate search errors because if the participants did not foveate the microcalcification signal, they would not click on the stimulus to activate the decision support. Interactive decision support is most helpful in mitigating decision and recognition errors, misses that occur when the observer foveates the signal but reports it as absent.

We omitted from displaying the probability of malignancy scores associated with each box because the per-pixel probability thresholds used to generate the connected components varied across the 4 CNN models. For instance, the probability of malignancy associated with cues for the microcalcification-3D stimuli would range between 0.9 and 1, whereas the probability scores would range between 0.2 and 1 for the microcalcification-2D stimuli. The difference in the range of probability scores across models would introduce information to the observer, potentially confounding our analysis.

To convert the connected components in a 3D phantom stimulus into bounding boxes overlaid on the stimulus, we first computed each component's center-of-mass coordinate (x, y, z). We then computed the Euclidean distance in the (x, y) plane between every pair of center-of-mass coordinates. We also computed the Manhattan distance between the z-coordinates for every pair of center-of-mass coordinates. We grouped components if their Euclidean distance was less than a model-specific distance threshold and their Manhattan distance was less than a model-specific threshold. Each group of connected components was converted into a single cue. The cue was placed at the mean location of all the center-of-mass coordinates in the group. The same procedure was done for the 2D connected components without the Manhattan distance calculation. The threshold parameters for this process can be found in Table 1, columns 5 and 6.

This grouping procedure was done to prevent overlap amongst the cued locations, which would induce visual clutter and distract from the primary task. Second, we wanted to reduce the average number of visual prompts per stimulus to less than 10 to maintain consistency with previous studies that report the number of CNN-CADe false positive prompts per image (see Table 3 in [39]). Third, we attempted to normalize the number of cues in the signal-present and signal-absent sets of stimuli to prevent observers from utilizing this information to make their decisions. For example, in the edge case where all signal-present stimuli have at least one cue and all signal-absent stimuli have zero cues, a human observer could use the number of cues on the stimulus to determine the presence/absence of the signal.

Lastly, we equated the localization accuracy of the 4 models to ensure, from the observer's vantage point, that the CNN-CADe provided consistent, accurate information across the 2D/3D searches for the mass and microcalcification signals. The localization accuracy in 3D was defined as the proportion of signal-present trials where at least one cue prompt was less than 1.5 dva away from the centroid of the signal profile in the (x, y) plane. Moreover, the central slice of the signal needed to appear in at least one of the slices where the cue would appear on the screen. Recall that the cue boxes spanned 5 slices in z for the microcalcification signal and 11 slices in z for the mass signal. For 2D, only the former condition described for 3D needed to be met to define CNN-CADe localization accuracy.

To equate the localization accuracy across the 4 models, we applied a grid search over 3 parameters in 3D: the P(malignancy) threshold, the Euclidean distance threshold, and the Manhattan distance threshold. For the 2D models, we applied a grid search over only the first two parameters. This grid search produced a localization accuracy of 0.7775 across all models (Table 1, column 3). In other words, on 31 of the 40 signal-present trials, at least one cue would be placed directly over the signal.

### 2.7. Human performance measures and statistical analysis

We assessed overall human performance in the search tasks based on the following primary endpoints: AUC, hit rate, and false alarm rate. We supplemented this analysis by stratifying misses into two categories: search errors and recognition errors. Furthermore, we quantified the proportion of the search area observers explored with eye movements (proportion of area covered by the Useful Field of View, or PAC UFOV) and the time participants spent searching. Lastly, for the peripheral detection task, we calculated the AUC for each signal to determine how much the mass signal was more detectable in the visual periphery than the microcalcification signal. We also combined the peripheral detectability measurements into participant-specific UFOVs (PUFOVs) to highlight individual differences for the CNN-CADe benefit in the search tasks. Below, we provide a more complete description of each analysis.

### 2.7.1. AUC-Search

We employed a multi-reader multi-case (MRMC) analysis [40–42] to evaluate significant differences in the AUC with versus without CNN-CADe decision support using the open-source MRMCaov software package available in the R programming language [43]. This software treats "readers" and "cases" as random effects under a generalized linear mixed effects model framework. The software also provides individual AUC estimates for each participant in each condition (with and without the CAD), assuming a binormal model. We applied this analysis 4 times, once for each search task: microcalcification 2D search, microcalcification 3D search, mass 2D search, and mass 3D search. We applied the Benjamini-Hochberg false discovery rate (FDR) correction to an $\alpha = 0.05$ level for all 4 two-tailed p-values [44].

In line with our primary hypotheses outlined in the introduction, we assessed whether the benefit of the CNN-CADe was significantly greater in 3D than in 2D for the microcalcification signal. For the mass signal, we determined whether this benefit was significantly greater in 2D than in 3D. Here, we utilized a nonparametric bootstrap resampling procedure (i.e., sampling readers and cases with replacement 20,000 times) and computed the mean empirical AUC for all 8 levels of the 3 within-subject factors. For a given bootstrap iteration and signal type, we subtracted the mean AUC without the CNN-CADe from the mean AUC with the CNN-CADe for both the 2D and 3D searches. For the microcalcification signal, we subtracted the difference in AUC in 2D from the difference in AUC in 3D. For the mass signal, we subtracted the difference in AUC in 3D from the difference in AUC in 2D. We computed the proportion of difference of differences in mean AUC that were greater than 0 across all 20,000 bootstrap iterations to obtain 1-tailed p-values. In total, two p-values, one for each signal, were compared to $\alpha = 0.05$.

### 2.7.2. Hit rate and false alarm rate

Hits and false alarms were defined as ratings greater than or equal to 5 on signal-present and signal-absent trials, respectively. The number of hits divided by the number of signal-present trials (40) produced a participant-specific hit rate. The same procedure was applied to false alarms on signal-absent trials. We utilized the bootstrapping procedure discussed above (i.e., sampling readers and cases with replacement 20,000 times) to obtain differences in mean hit rate or false alarm rate for the searches with the CNN-CADe and without it. The count of bootstrapped differences in the hit rate (or false alarm rate) more extreme than 0 was divided by 20,000 and then multiplied by 2 to obtain a two-tailed p-value. We FDR corrected for 4 p-values (2D microcalcification, 2D mass, etc.) per endpoint. This nonparametric procedure, including the number of pairwise comparisons and the FDR correction, was applied to search and recognition errors, the area covered by the UFOV, and the amount of time spent searching.

### 2.7.3. Search and recognition errors

Search and recognition errors allowed us to ascertain the impact of foveal vision on detection performance by stratifying misses into two distinct categories. Search errors were defined as the subset of false negative responses where an observer failed to fixate directly on the signal. Recognition errors were defined as the complement set of misses where observers missed the signal but stared directly at it [6,11,18]. In the 2D search conditions, we computed the Euclidean distance between every recorded fixation position and the center (x, y) coordinate of the signal's location for a given participant and trial. If at least one fixation was at a distance less than or equal to 2.5 dva away from the signal, then the observer fixated the signal on that trial. For the 3D search conditions, we augmented the definition of fixating the signal because its profile spanned multiple consecutive slices. The Manhattan distance between the z coordinate of every fixation and the signal's central slice z coordinate was computed. If the Manhattan distance for a fixation was less than or equal to N, where N=3 for the microcalcification and N=10 for the mass, and the Euclidean distance in (x, y) was less than or equal to 2.5 dva, then that fixation was considered to be on the signal in 3D. To obtain an error rate per participant, we divided the count of each type of error by the total number of signal-present trials (40).

**2.7.4. PAC UFOV**

The proportion of the search area covered by the UFOV provides an approximate estimate of how much observers explored the 2D slices or 3D volumes with eye movements. For a given observer and trial, we "painted" a circle on all recorded (x, y) fixation locations in 2D and all recorded (x, y, z) fixation locations in 3D. The circle had a radius of 2.5 dva, the standard in the literature [6,18]. (We include a supplementary analysis that utilizes a signal-specific UFOV radius based on each observer's peripheral detectability of a signal). We computed the cardinality of the union set of pixels that were "painted" by the UFOV and divided this count by the number of pixels that comprised the DBT phantom slice (2D) or DBT phantom volume (3D) to obtain a proportion. We computed point estimates per observer by averaging the PAC UFOV across all signal-present and signal-absent trials.

**2.7.5. Search time**

The search time was defined as the elapsed time (in seconds) between when the image stimulus was first displayed on the monitor and when the participant pressed the spacebar to end the search component of the trial. Point estimates were obtained by averaging across all signal-present and all signal-absent trials.

**2.7.6. AUC-Peripheral detectability**

We implemented a single-reader multi-case analysis to obtain participant-specific AUCs for the microcalcification and mass signals based on their rating data from the peripheral detection task. This was done using the MRMCaov software package available in R. To test whether the average AUC, across participants, was significantly lower for the microcalcification signal than the mass signal, we utilized the same nonparametric bootstrapping procedure (20,000 bootstraps) with empirical AUCs to obtain differences in mean AUCs across the two signals. A single p-value was compared to $\alpha = 0.05$. One participant chose not to complete the peripheral detection task for the mass signal, and we omitted them from this analysis.

Lastly, we used the parametric AUC estimates from the MRMCaov package to obtain participant-specific UFOV radii for each signal type (PUFOV). We assumed that detection performance for both signals at the fovea in a location-known-exactly and signal-known-statistically task would produce AUC estimates of 1, which is a mild assumption. We then fit a half-Gaussian function:

$$AUC = \frac{\gamma}{\sqrt{2\pi\sigma^2}} e^{\frac{-E^2}{2\sigma^2}}$$

where $\gamma$ and $\sigma^2$ are fitting parameters and E refers to eccentricity ($0 \leq E \leq 10$ degrees visual angle). The function was fit to two points for a single signal: ($E = 0$, $AUC = 1$) and ($E = 5$, $AUC = \widehat{AUC}$) where $\widehat{AUC}$ refers to the participant-specific estimated AUC from the MRMC model described above. We set $AUC = 0.82$ and solved for $E$ to obtain $E^*$. The radius of the PUFOV for a given signal was set to $E^*$. An example of applying this procedure to a single subject is shown on the left-hand side of Figure 4a/b for the microcalcification and mass signals, respectively. We acknowledge that we are fitting a function with two parameters to 2 data points; thus, our fit has no error. This limitation can be fixed in future work by computing the peripheral detectability of each signal at various eccentricities.

We recomputed the mean proportion of area covered on 2D and 3D signal-absent trials without the CNN-CADe available, using each participant's PUFOV. The area covered in signal-absent trials without the CNN-CADe provides a measure of eye movement exploration that is not confounded by finding the signal during the search or relying on the CNN-CADe prompts. The personalized UFOV, instead of the standard UFOV, normalizes peripheral detectability for a given signal across all participants by combining the two constructs into one. Next, we correlated these estimates of the PAC PUFOV with each observer's change in AUC when searching with the CNN versus without it. This correlation allowed us to ascertain whether those who explored less, while normalizing by their peripheral detectability of each signal, benefited the most from the CNN-CADe during the searches. That is, they have the most considerable change in AUC.

## 3. RESULTS

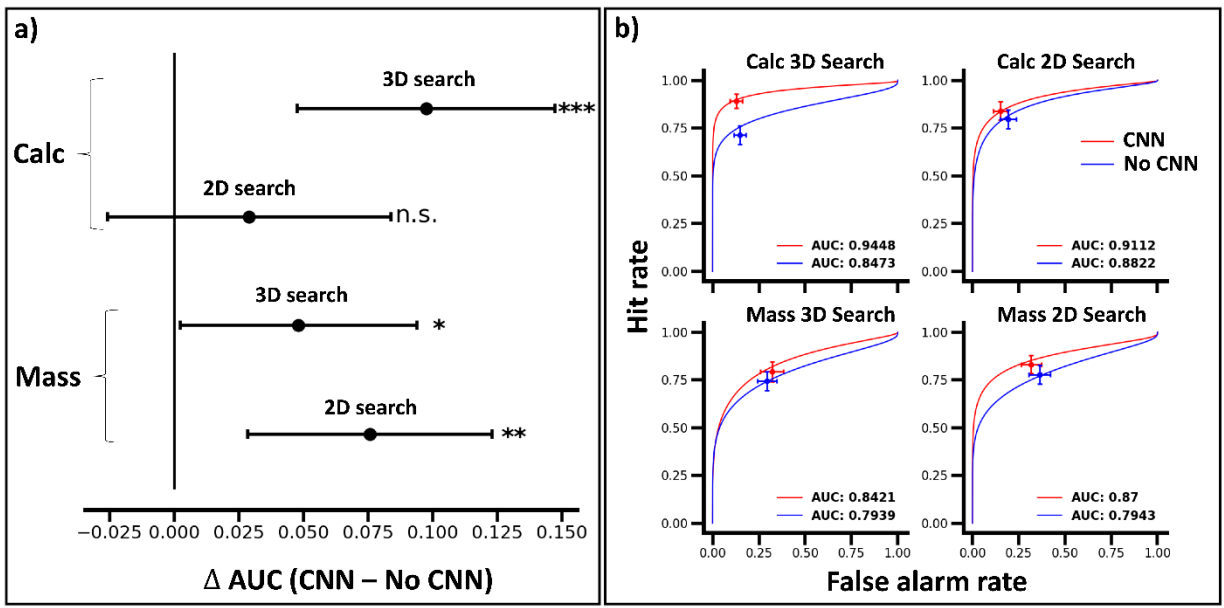

Figure 2. The results of the MRMC analysis depict the benefit of the CNN-CADe during the 2D and 3D searches for the microcalcification and mass signals. a) The difference in reader-averaged AUC between the CNN and no CNN searches (scatter points) and their respective 95% confidence intervals (horizontal lines) are plotted with respect to the null hypothesis of no change in AUC when searching with versus without the CNN-CADe (vertical line centered at 0). From top to bottom, the change in AUC is plotted for microcalcification 3D search, microcalcification 2D search, mass 3D search, and mass 2D search. * = $p < 0.05$, ** = $p < 0.01$, *** = $p < 0.001$ and n.s. signifies $p > 0.05$. b) reader-averaged Binormal ROC curves with (red lines) and without the CNN-CADe (blue lines) are shown for microcalcification 3D search (top left), microcalcification 2D search (top right), mass 3D search (bottom left) and mass 2D search (bottom right). The area under the ROC curve is also reported for the CNN and No CNN searches in each of the 4 subplots. The scatter point in a given subplot represents the participant-averaged operating point (false alarm rate, hit rate) at the cut point 4.5, the middle of the rating scale used in our experiment. Horizontal and vertical error bars for a given operating point denote 68% bootstrapped confidence intervals (~1 standard error of the mean) for the false alarm rate and hit rate, respectively.

### 3.1. The CNN-CADe provides the largest benefit for the 3D search of small signals

Our main objective in this study was to determine if the CNN-CADe improves performance in the 2D and 3D searches and if the benefits are contingent on what type of signal observers had to find. Figure 2a demonstrates that the change in overall search performance with the CNN-CADe in 2D and 3D depended on the signal type. For the small microcalcification signal, having the CNN-CADe available during the 3D search markedly improved the overall AUC ($\Delta AUC = 0.098$, 95% CI $[0.048, 0.147]$, $p = 0.0002$). However, in the 2D search, the observed change in AUC did not reach statistical significance ($\Delta AUC = $ 0.029, 95% CI $[-0.026, 0.084]$, $p = 0.296$). Moreover, the benefit of the CNN-CADe in 3D was significantly greater than that of the CNN-CADe in 2D ($\Delta\Delta\,AUC = 0.066$, $p = 0.035$).

For the mass signal, we observed an opposite effect of the CNN-CADe on 2D versus 3D search performance. During the 3D search for the mass, the AUC change was significant but did not survive an FDR correction ($\Delta\,AUC = 0.048$, 95% CI $[0.002, 0.094]$, $p = 0.048$). On the other hand, when observers searched for the mass signal in 2D, the CNN-CADe significantly benefited their search ($\Delta\,AUC = 0.076$, 95% CI $[0.028, 0.123]$, $p = 0.002$). However, the improvement in AUC when searching with the CNN-CADe in 2D was not significantly greater than the improvement in AUC when searching for the mass in 3D ($\Delta\Delta\,AUC = 0.033$, $p = 0.133$).

### 3.2. The influence of the CNN-CADe on hit and false positive rates

We evaluated criterion-specific search performance measures to understand further how the CNN-CADe influenced the observer's perceptual decision-making processes. Figure 2b depicts the reader-averaged ROC curves and the mean operating points (i.e., false alarm rate and hit rate pair) at the rating threshold 4.5 when observers searched with and without the CNN-CADe for the two signals in 2D and 3D.

When observers searched for the microcalcification in 3D (Figure 2b, top left), the average hit rate significantly increased from 0.714 to 0.892 when the additional information from the CADe was made available to them, $p = 0.001$. We observed a modest but not significant reduction in the mean false alarm rate as well ($FAR_{No\ CNN} = 0.146$, $FAR_{CNN} = 0.127$, $p = 0.6844$). Similarly, when searching for the microcalcification in 2D (Figure 2b, top right), the CNN-CADe increased the mean hit rate from 0.797 to 0.840, and reduced the average false alarm rate from 0.192 to 0.150. However, neither the difference in hit rate nor the difference in false alarm rate was statistically significant from 0, $p = 0.384, p = 0.317$, respectively. Together, these results suggest that CNN-CADe facilitated the detection of the microcalcification in 3D to a greater extent than in 2D and support the finding of a significantly larger change in AUC in 3D than in 2D, as discussed above.

In considering the search for the mass signal, the CNN-CADe minimally impacted the hit and false alarm rates in both the 2D and 3D modalities. During the 3D search (Figure 2b, bottom left), the mean hit rate increased with the CNN-CADe ($HR_{No\ CNN} = 0.745$, $HR_{CNN} = 0.794$) but so did the false alarm rate ($FAR_{No\ CNN} = 0.293$, $FAR_{CNN} = 0.320$). However, these differences in the hit rate and false alarm rate were not significantly different from 0 ($p = 0.135$, $p = 0.624$, respectively). When observers searched for the mass signal in 2D, their average hit rate increased from 0.778 to 0.831 with the CNN-CADe, but this difference was not significantly different from 0 ($p = 0.130$). The false alarm rates were also not significantly different from one another ($p = 0.368$) despite the CNN-CADe marginally reducing the mean false alarm rate from 0.363 to 0.317. The negligible changes in hit rate and false alarm rates support the finding that the improvement in AUC in the 2D search was not significantly higher than the improvement in AUC in the 3D search for the mass signal.

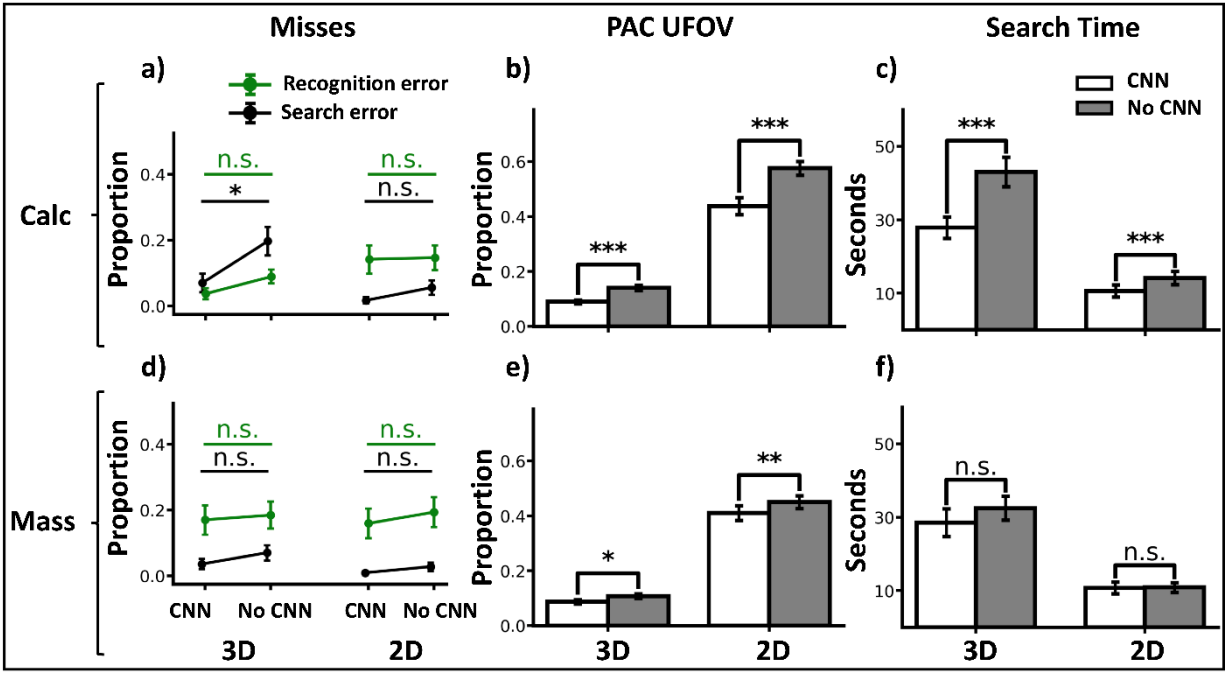

Figure 3. Additional measures exemplify the impact of the CNN-CADe on the participant's search strategies. a) The mean proportions of microcalcification search errors (black lines) and recognition errors (green lines) in 3D (left set of lines) and 2D (right set of lines). For a given line, the left scatter point represents a particular mean error rate when observers searched for the microcalcification with the CNN-CADe, and the right scatter point represents that same endpoint but when observers searched for the microcalcification without the CADe available. b) The mean proportion of the search area (PAC) covered by the standard UFOV (2.5 dva radius) on all trials for the 3D (left cluster of bars) and 2D (right cluster of bars) searches. White bars denote the area covered when searching for the microcalcification signal with the CNN-CADe, and gray bars represent the proportion of the area covered when searching for the microcalcification without the CNN-CADe. c) The mean search time while looking for the microcalcification signal. The breakdown of search time in 2D/3D and CNN/No CNN is kept consistent with b). d), e), and f), The same endpoints and organization of 2D/3D and CNN/No CNN data that was discussed in a), b), and c) are shown for the mass signal. All error bars represent 68% bootstrap confidence intervals (~ 1 SEM). * = p < 0.05, ** = p < 0.01, *** = p < 0.001 and n.s. signifies p > 0.05.

### 3.3. The influence of the CNN-CADe on search and recognition errors

We are not only interested in whether the CNN-CADe improves search performance in 2D and 3D for both small and large signals but also in how it facilitates the detection of these signals. Our analysis of the gaze-contingent errors provides this additional insight. Figure 3a depicts how search and recognition errors for the microcalcification signal in 3D and 2D changed when the bounding boxes from the CNN-CADe were made available during the searches. Figure 3a left demonstrates that the mean search error rate (SER) was significantly reduced with the CNN-CADe ($SER_{No\ CNN} = 0.197$, $SER_{CNN} = 0.070$, $p = 0.0069$). Although the mean recognition error rate (RER) was also significantly reduced from 0.089 to 0.038 when searching with the CADe, $p = 0.043$, it did not survive an FDR correction. When observers searched for the microcalcification signal in 2D (Figure 3a, right), we observed that the CADe reduced the mean SER from 0.056 to 0.017 and the mean RER from 0.147 to 0.142. However, these differences were not significantly different from 0, $p = 0.060$ and $p = 0.893$, respectively.

Relative to Figure 3a, Figure 3d shows similar but less dramatic effects of the CNN-CADe on the search and recognition errors for the mass signal in both 3D and 2D. During the 3D search (Figure 3d, left), both the mean search error rate ($SER_{No\ CNN} = 0.071$, $SER_{CNN} = 0.036, p = 0.125$) and mean recognition error rate ($RER_{No\ CNN} = 0.185$, $RER_{CNN} = 0.170, p = 0.647$) were reduced, but the changes in error rates were not significantly different from 0. When observers searched for the mass signal in 2D (Figure 3d, right), the mean SER was reduced from 0.028 to 0.009, and the mean RER was reduced from 0.1938 to 0.160 when the cue boxes were present during the search. However, these differences in SER and RER were not significantly different from 0 ($p = 0.139$, $p = 0.305$, respectively).

### 3.4. CNN-CADe reduces eye movement exploration (PAC UFOV)

Figure 3b summarizes the 2D/3D PAC UFOV when participants were tasked to look for the microcalcification signal with and without the CNN-CADe. In the the 3D search (Figure 3b, left), the mean PAC without the CNN-CADe (0.140) was significantly higher than the mean PAC with the CNN-CAD (0.090), $p < 5e^{-5}$. Similarly, during the 2D search (Figure 3b, right), the PAC without the CNN-CADe available (0.576) was significantly higher than with it (0.438), $p < 5e^{-5}$.

We observed a similar trend in the PAC UFOV when observers were tasked to find the mass signal with and without the CNN-CADe in both the 2D and 3D searches (Figure 3e). The PAC during 3D search (Figure 3e, left) was significantly lower when searching with the CAD (0.087) than without it (0.108), $p = 0.013$. Figure 3e, right, shows that the PAC in 2D was also significantly lower with the CADe (0.410) as opposed to searching without it (0.450), $p = 0.003$. In sum, regardless of the imaging modality or signal type, observers, on average, explored less of the search area with eye movements when the CNN-CADe support system was enabled.

### 3.5. CNN-CADe reduces the search time for the microcalcification but not the mass signal

Figure 3c depicts the effect of the CNN-CADe on the time spent searching for the microcalcification in 3D and 2D. During the 3D search (Figure 3c, left), there was a significant reduction in average search time from 42.994 seconds without the CADe to 27.848 seconds with it, $p < 5e^{-5}$. While performing the 2D search (Figure 3c, right), observers searched for 14.141 seconds on average without the CADe and 10.606 seconds with the CADe, and this difference was statistically significant, $p < 5e^{-5}$.

Figure 3f exemplifies how the CNN-CADe impacted the search time for the mass in both 2D and 3D. Figure 3f, left shows a marginal reduction in search time when observers looked for the mass in 3D with the CNN-CADe (28.503 seconds) versus without it (32.445 seconds), $p = 0.226$. We also observed a slight reduction in 2D search time with the CNN-CADe (10.751 seconds) versus without it (10.887 seconds), $p = 0.842$. However, neither the differences in 3D search time nor 2D search time were statistically significant.

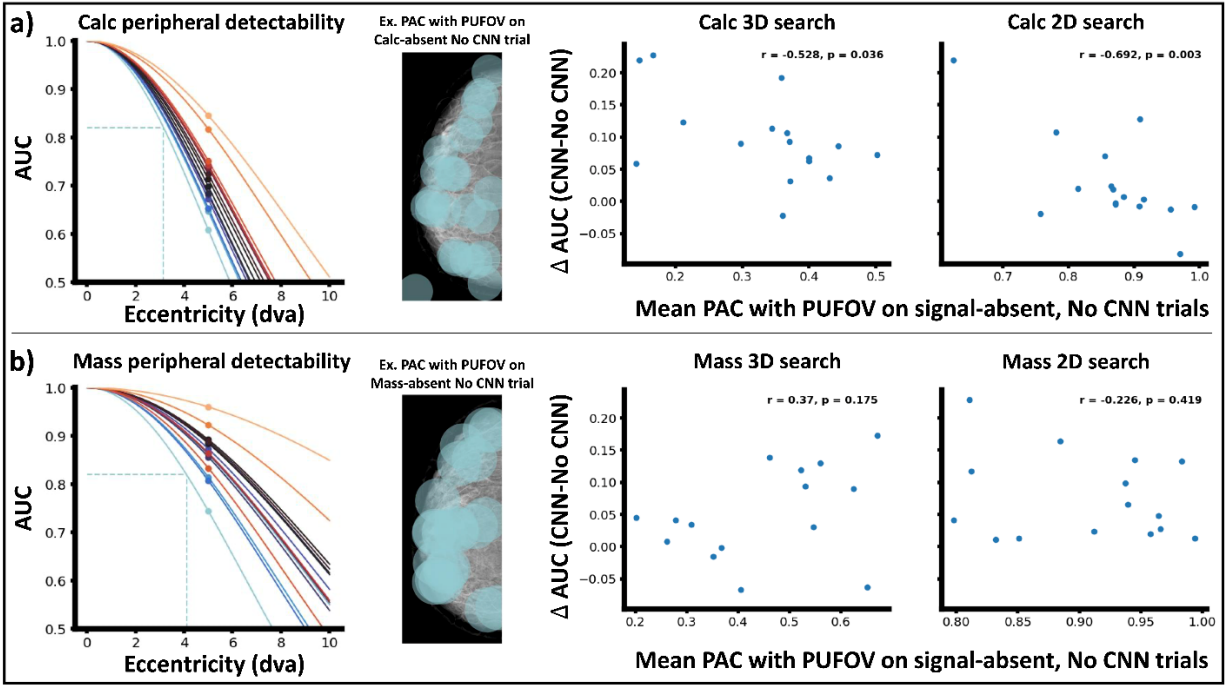

Figure 4. The procedure for deriving the PUFOV for each signal/participant combination. The correlation between the change in overall performance versus the mean PAC with the PUFOV on signal-absent, no CNN trials for the mass, and microcalcification signals in both 2D and 3D are also shown. a) Left, a half Gaussian is individually fit to each participant's AUC in the microcalcification peripheral detection task. Each colored line represents a fit for a different participant. Scatter points represent each participant's AUC from the forced-fixation yes/no detection task for the microcalcification signal presented at 5 dva away from the fixation point. The blue dotted horizontal line intersects the y-axis at 0.82. The blue vertical line represents the predicted eccentricity at which that participant would have an AUC of 0.82 in the peripheral detection task. This eccentricity served as the radius of the circular PUFOV for the microcalcification signal for that participant. a) Middle right, the PUFOV for the microcalcification signal derived in 4a, left is "painted" on all recorded fixation positions from an example 2D microcalcification signal-absent search trial without the CNN-CADe cues available to the participant. Blue circles overlaid on the DBT slice visualize the 2D area covered by the PUFOV. a) Middle right, scatterplot relating the change in AUC (CNN AUC – No CNN AUC) during the microcalcification 3D search versus the PAC PUFOV on signal-absent microcalcification 3D trials without the CNN-CAD output available. Each point represents a single participant's change in AUC (y-axis) and how much they explored with eye movements while accounting for their peripheral detectability of the microcalcification signal (x-axis). The correlation test statistic, $r$, and the corresponding p-value are included in the subplot legend. a) Right, the dependent variables represented in the scatterplot in a) Middle right are visualized for the microcalcification 2D search task. b) left, middle left, middle right, and right depict the same analysis as in a) but for the mass signal.

### 3.6. The mass signal is more detectable in the visual periphery than the microcalcification

The scatter points in Figure 4a, left, and Figure 4b, left plot the peripheral detectability of the microcalcification and mass signals for all participants in this study, respectively. The mean AUC for detecting the microcalcification in the visual periphery was 0.711. In contrast, the mean AUC for detecting the mass in the visual periphery was 0.863, and this difference was statistically significant ($p < 5e^{-5}$).

### 3.7. Deriving signal-specific UFOVS for each observer (PUFOVS)

Figure 4a, left, also includes the half-gaussian fits to each observer's peripheral detectability of the microcalcification signal. Figure 4b, left, shows the same type of fits but for the mass signal. The half-gaussian fits were used to derive signal-specific UFOVs (PUFOVs) for each participant at an AUC threshold of 0.82. The middle left subplots of Figure 4a and Figure 4b depict the process of "painting" the PUFOVs on all recorded fixation positions from a single participant during the 2D searches of the microcalcification and mass signals, respectively. These fixations were obtained from two searches on signal-absent trials without the CNN-CADe output available. Of note for the participant's data shown here is that the radius of the circular PUFOV for the mass signal is larger than the radius of the PUFOV for the microcalcification signal. Thus, when incorporating the PUFOV into the computation of the PAC, more of the DBT phantom slice is covered in the mass trial than in the microcalcification trial.

### 3.8. Individual differences in the AUC benefits of the CNN-CADe correlate with PAC PUFOV

Our last analysis evaluates whether those who explore less in 2D/3D with their PUFOV benefit the most from the CNN-CADe. Figure 4a, middle right exemplifies this relationship for the microcalcification 3D search. We observed a strong negative linear correlation between how much people explore with eye movements and their change in AUC when searching with the CADe versus without it ($r = -0.528, p = 0.036$). In short, those who tended to explore less of the 3D volume with eye movements benefited the most from the CNN-CADe.

The correlation analysis reported above depends on two free parameters: 1) the half-gaussian function we chose to fit the peripheral detectability data to, and 2) the AUC threshold of 0.82 for computing the radius of the PUFOVs for each observer. Therefore, we reran the analysis using AUC thresholds ranging from 0.82-0.9 in steps of 0.02. We also fit a line rather than a half-gaussian to the peripheral detectability estimates and used the same range of AUC thresholds. Across the set of AUC thresholds and the two fitting functions (10 models), the correlations ranged from $-0.528$ to $-0.416$ (mean $= -0.4793$, std $= 0.0463$).

We also ran this same analysis for the 2D search of the microcalcification (Figure 4a, right). Like the microcalcification 3D search, we observed a strong negative linear relationship between these two variables ($r = -0.692, p = 0.003$). Furthermore, considering the range of AUC thresholds and two fitting functions, the correlations spanned from -0.692 to -0.607 (mean = -0.6624, std equals $.0286$). However, caution should be taken in interpreting the strength of this negative linear relationship because one participant was an outlier (Figure 4a, right, top left corner). Removing this person from the analysis produced a correlation of $-0.373$ ($p = 0.171$) at an AUC threshold of 0.82 for the half-gaussian fitting function.

We ran the same analysis for the mass signal in the 2D and 3D searches to understand if this relationship holds across both small and large signals. For the mass 3D search (Figure 4b, middle right), we observed a positive linear relationship between the mean PAC with the PUFOV on signal-absent trials without the CNN-CADe and the change in AUC between searching with versus without the CNN-CADe. However, this relationship was not statistically significant ($r = 0.370, p = 0.175$). Correlations ranged from 0.185 to 0.448 (mean $= 0.3302$, std $= 0.075$). For the mass 2D search (Figure 4b, right), we observed a negative linear relationship between these two dependent variables, but this relationship was also not significant ($r = -0.226, p = 0.419$). Here, the correlations ranged from $-0.226$ to $-0.114$ (mean $= -0.169$, std $= 0.047$).

## 4. DISCUSSION

Our main objectives in this experiment were to assess 1) how the benefits of the CNN-CADe vary across 2D and 3D searches and 2) how the support system interacts with the size of the searched signal across these two imaging modalities. Our results show that the CNN-CADe brings about added benefits for the 3D search of small signals, and to a lesser extent, it improves the 2D search of large signals. To better understand this nuanced interaction, we quantified how much the CNN-CADe mitigated the microcalcification and mass search and recognition errors across the 2D and 3D modalities.

For example, at the outset, we hypothesized that the CNN-CADe provides unique benefits to the 3D search of the small microcalcification signal by guiding an observer's eye movements to suspicious locations cued by the model observer, effectively reducing search errors. Recall search errors result from the interaction between under-exploring the 3D volumetric data with eye movements and having low detectability of the microcalcification signal in the visual periphery [9,28,45,46]. Conversely, when searching for the microcalcification signal in 2D, it is relatively easy to direct one's center of gaze to most regions of the DBT slice in a time-efficient manner. Thus, we predicted that the benefit of the CNN-CADe would be less pronounced in this case because extra-foveal processing would have a diminished influence on search errors. Indeed, our results show that the CNN-CADe markedly reduced search errors in 3D but not in 2D (Figure 3a). These results are commensurate with the fact that the CNN-CADe induced a significant increase in 3D search AUC but only a marginal improvement in 2D search AUC (Figure 2a). Moreover, the difference in AUC for the 3D search was significantly higher than in the 2D search.

We can ascertain that the participants relied heavily on the marked locations made by the CNN-CADe when searching for the microcalcification signal because not only did the search error rate decrease when the model output was available, but participants, on average, explored less with eye movements (Figure 3b). They also searched for a shorter period when the cued locations were available (Figure 3c). Thus, if an observer adopts a search strategy focusing on visually inspecting the cued locations and the cues are highly accurate, we expect observers to explore less and for a shorter duration while maintaining high sensitivity and specificity [47]. These findings highlight the importance of having an accurate auxiliary aid when performing life-critical tasks such as early cancer screening. Prior work has shown that inaccurate CADe systems can increase misses because observers explore less of the search space with eye movements when it is made available to them [48], a possible consequence of *automation bias* or overreliance on the machine [49]. This effect can be particularly pernicious as 3D imaging modalities become the standard of care for breast cancer detection.

Our second hypothesis posited that the CNN-CADe would benefit the detection of the mass signal in 2D to a greater extent than in 3D. The mass signal is more detectable in the visual periphery than the microcalcification signal (Figure 4a, left versus Figure 4b, left), and it spans many more slices in 3D than the microcalcification signal. Thus, more signal information in 3D can be integrated. However, in 2D, the simulated glandular and adipose tissue in the DBT slice can obfuscate or visually mask the mass signal [50,51]. The signal profile is dominated by low spatial frequency information, and there is high energy at low spatial frequencies in the noise power spectrum of the DBT phantom slice [25]. In sum, we expect more recognition errors and false positives in 2D than in 3D, and the CNN-CADe should mitigate these 2D errors.

Our results partially align with this hypothesis because the CNN-CADe improved the overall search AUC for the mass in 2D but not in 3D (Figure 2a). However, the improvement in 2D was not significantly greater than in 3D. Despite observing a significant improvement in AUC for the 2D search with the CNN-CADe, we did not find a significant reduction in false alarms (Figure 2b, bottom right) or recognition errors (Figure 3d, right). Interestingly, observers explored less of the 2D DBT slice with the CADe available (Figure 3e) but did not spend significantly less time searching (Figure 3f). One interpretation is that in the presence of a "second opinion," participants spent more time scrutinizing only the cued locations.

An important finding from this work is the considerable inter-observer variability in the benefits of the CNN-CADe on 3D search for small signals. Moreover, this variability can be related to an observer's exploration behaviors, quantified via the PUFOV. This claim is realized by our analysis of individual differences (Figure 4a, middle right), which demonstrates a negative correlation between the change in AUC when searching with the CADe (versus without it) and the mean proportion of the search area covered with the PUFOV on signal-absent trials with no CNN-CADe. This finding suggests that those who explored less of the 3D DBT phantom benefited the most from the cued locations. Additionally, our PUFOV construct considers the peripheral detectability of the microcalcification signal. Therefore, if an observer makes many eye movements during the search but has poor peripheral detectability, they should still benefit from the CNN-CADe because their poor peripheral vision will reduce their effective eye movement exploration.

On the other hand, we did not observe a significant correlation between the PAC with the PUFOV and the change in AUC when participants searched for the mass signal in 2D or 3D (Figure 4b, middle right and right). The detection of the mass signal, with signal location uncertainty, is noise-limited [52]. That is, observers should not be influenced by extra-foveal processing and eye movement exploration but rather by signal contrast and how this attribute interacts with the anatomical noise embedded in the DBT phantom. In this regard, properly placed CADe prompts on suspicious locations scrutinized by the observer may induce increased confidence in their decision because the model reassures the observer's initial suspicion.

Our study has inherent limitations worth enumerating to help contextualize our results within the broader medical imaging field. First, we utilized trained human observers as opposed to radiologists. Given the data-intensive nature of our experiment, we opted to run non-expert observers because it allowed us to run many trials while collecting eye-tracking data. As a consequence, the external validity of our findings may be limited in scope because expertise mediates observer performance [53,54]. Despite differences in performance due to expertise, studies have shown how bottlenecks and properties of the visual system common to naïve and radiologist observers [55] result in similar effects across the two cohorts [9,56,57]. For instance, [9] demonstrated that trained human observers and radiologists are similarly susceptible to making search errors when tasked to find microcalcification-like signals in 3D volumetric images. This finding can be explained by the neurophysiological constraints of the human visual system rather than expertise. Hence, we would expect a CNN-CADe system to aid the detection of microcalcifications in DBT volumes within a clinical setting.

Another limitation of our study concerns how our 3D search task differs from how 3D volumetric images are interpreted in a clinical setting. When radiologists interpret DBT data, they have available either a 2D mammogram or a 2D synthetic (2D-S) view generated from the DBT data. Prior work has demonstrated how a 2D-S can guide eye movements in 3D and thus mitigate search errors [58]. Therefore, the presentation of CADe prompts may provide redundant information that can otherwise be extrapolated from the 2D-S. The CNN-CADe may only provide radiologists with little additional benefit beyond a reduction in reading time [59].

The signals used in our study pose additional caps on the external validity of our findings. First, observers knew which signal to search for at the outset of every trial. However, radiologists are not privy to this information in practice—there is signal uncertainty. Therefore, radiologists must maintain multiple signal templates in memory while examining medical images. Second, microcalcifications often appear in clusters. However, here, we had observers search for a single microcalcification. However, prior work has shown that microcalcification clusters have low peripheral detectability [29]. Our results suggest that the benefits of the CNN-CADe in 3D search may extend to microcalcification clusters. Third, there are signals radiologists screen for indicative of malignant lesions, particularly architectural distortions, that were not investigated in this study. Despite the differences between our experimental design and what is observed in clinical practice, future studies with radiologists (or radiology residents) can measure their peripheral detectability of various signals and quantify how much they explore 3D volumetric images with eye movements. Our work provides specific predictions about how a CNN-CADe would benefit the 3D search with those measurements in hand.

Lastly, cognitive factors such as fatigue [60] and criterion shifts that arise from low target prevalence rates in cancer screenings [61] might interact with the search effects in this study. Similarly, our study included one signal per case, not addressing instances with multiple lesions, often leading to the satisfaction of search [62–64].

# 5. CONCLUSION

Recent advances in artificial intelligence-based computer-aided detection algorithms can improve human observer search performance in 3D volumetric medical images where interpretation time and effort far exceed the visual examination of 2D medical images. Our study suggests that CNN-CADe brings about greater performance benefits to the 3D search of small signals (vs. 2D search) by reducing search errors caused by the under-exploration of the volumetric data. Our proposed methodology for measuring observer 3D search under exploration has the potential to identify individuals who would benefit the most from the CNN-CADe support system.

## Disclosures

The authors have no conflicts of interest to report that would influence the conclusions drawn in this paper.

## Code and data availability

The human observer data for this paper can be found at https://doi.org/10.17605/OSF.IO/CVA6K.

## Acknowledgments

This work was supported by the National Institute of Health under Grant R01EB026427 and sponsored by the U.S. Army Research Office. It was accomplished under Contract Number W911NF-19-D-0001 for the Institute for Collaborative Biotechnologies. The views and conclusions contained in this document are those of the authors. They should not be interpreted as representing the official policies, either expressed or implied, of the U.S. Government. The U.S. Government is authorized to reproduce and distribute reprints for government purposes, notwithstanding any copyright notation herein.